# Dissecting Embedding Bag Performance in DLRM Inference

Chandrish Ambati   Jing Ding   Trung Diep

Celestial AI

Abstract

As the size of DLRMs gets larger, the models must be partitioned across multiple GPUs or nodes of GPUs due to the size limitation of total HBM memory that can be packaged in a GPU. This partitioning adds communication and synchronization overhead of sending and receiving data across GPUs. We use the NCCL and NVSHMEM libraries to measure the performance of an Embedding Bag kernel implemented on H100 GPUs. We compare its performance across different batch sizes, number of tables, table sizes, pooling factors, and embedding dimensions. For a large embedding table that spans multiple GPUs, we project the performance slowdown from distributing an embedding table across multiple GPUs.

## 1. Introduction

In recent years, the size of Deep Learning Recommendation Models (DLRM) has been increasing rapidly. We have seen from several billions to trillions of parameters requiring terabytes (TBs) of memory to store the model parameters. The latest data center GPUs like the Nvidia H100 GPUs [12] contain High Bandwidth Memory (HBM) in the order of tens of gigabytes (GBs). The available memory capacity is lower than the memory required for performing inference on these models. This gap has resulted in the need for employing distributed machine learning techniques in which the model parameters are distributed across multiple GPUs. There has been much research [3-6] that introduces splitting strategies for the models and how that affects the computational graph. With larger models, the communication and synchronization cost of sending and receiving data among a cluster of GPUs adds a significant overhead that incurs on top of the computation time of the model execution, despite many improvements in interconnect technology like Nvidia's NVLink [2] to improve interconnect latency and increase interconnect bandwidth.

On the software side, the programming model and algorithms used for communication primitives vary depending on the library used and the access patterns of sending and receiving data. We compare two such libraries, NCCL [5] and NVSHMEM [6], in the context of a Deep Learning Recommendation Model [2]. Recommendation models form the backbone of many websites and apps we use today like Amazon [7, 8], Netflix [9], Google [10], Meta [2], and TikTok [11] to name a few. The models are used for a variety of use cases like Click Through Rate (CTR) prediction, ranking of search results, and showing personalized feeds and advertisements. While the architecture of recommendation models used across companies can be different, one of the common elements found in each of these models is the usage of embeddings. Embedding is a dense vector representation of each of the levels of a categorical feature. The size of these embedding tables has grown by over 16 times [19] in recent times, with recent models having trillions of parameters. The growth of these embeddings has led to the increasing memory requirement of DLRM models. For DLRM models that are distributed across GPUs, the fraction of time spent on communication varies from 5% to 70% [18]. The communication and synchronization overheads contribute directly to how well or poorly the DLRM inference performs.

Optimizing DLRM performance, specifically the embedding pooling performance, is made difficult by the irregular memory accesses of the row retrievals of an embedding table. It is further complicated by the need for an entire embedding table to be distributed across multiple GPUs. One of the main goals of this study is to analyze the embedding pooling performance for various configurations with different pooling factors, batch sizes, number of tables, table sizes, and embedding dimensions. Another goal is to project the performance slowdown due to distributing the embeddings across multiple GPUs.

In this study, we make the following contributions:
  i) Provide a benchmarking framework on which DLRM performance can be approximated by distributing the embedding tables across multiple GPUs.
  ii) Implement an Embedding Bag kernel that can be used to compare the performance between NVSHMEM and NCCL implementations.
  iii) Analyze the difference between the NVSHMEM and NCCL libraries and compare the performance of commonly used communication primitives.

## 2. Background

Inter-Process communication (IPC) refers to the mechanisms provided by an operating system (OS) to enable data sharing between two processes. Message Passing communication between two processes is done by interchanging messages. Message Passing Interface (MPI) [23] library is one of the first and most popular implementations of message passing parallel model. Shared memory is another way two processes can share data, and OpenMP [25] is the first practical implementation of such a programming model. While shared memory programs are easier to develop, message passing usually achieves better portability and scalability across many nodes. Distributed Shared Memory (DSM) programming models aim to combine the advantages of both by supporting the notion of shared memory in a distributed architecture. Partitioned Global Address Space (PGAS) is a DSM model that provides global address space and a Single Program Multiple Data (SPMD) control model [24]. Examples of this implementation include Unified Parallel C (UPC), Co-Array Fortran (CAF), and Open-standard Symmetric Hierarchical MEMory (OpenSHMEM).

A heterogeneous system is a system in which one or more host CPUs are connected to one or more GPUs. Heterogeneous parallel programming models simplify these systems from a programming standpoint by differentiating between host (CPU) and devices (accelerators).

Open Computing Language (OpenCL) [26] is one such example, an open standard for general-purpose programming across CPUs, GPUs, and other processors. The Compute Unified Device Architecture (CUDA) programming model is widely used for writing machine learning or general-purpose applications on current Nvidia GPUs. CUDA software environment allows developers to use C++ as a high-level programming language in which parts of the code structured as CUDA kernels are executed on devices (GPUs) and other portions of the code are executed on the host (CPU).

In a distributed architecture, the patterns of communication can be divided into two major types: point-to-point and collective communication. In point-to-point, the communication is between two processes or devices, whereas in collective communication pattern, the data can be transferred to more than two devices.

Nvidia Collective Communication Library (NCCL) [6] is a library of bandwidth-optimized communication collectives for GPUs like all reduce, all gather, broadcast, and reduce scatter, among others. Widely used deep learning frameworks like PyTorch [21] and TensorFlow [22] use NCCL as their communication backend for GPUs. The NCCL implementation is optimized for high bandwidth on topology-aware interconnects, and its application programming interfaces (APIs) closely resemble that of MPI (Message Passing Interface) [23]. NVSHMEM is another multi-GPU programming library based on OpenSHMEM [22] that uses the PGAS memory model and can be accessed with finer-grained operations or one-sided read and write APIs.

Traditionally, CUDA kernels are used as a compute offload mechanism, in which compute-intensive portions of an application are written as CUDA kernels to be run on the GPUs whereas other parts of the program continue executing on the host. The NCCL APIs which facilitate communication among GPUs are called from the host side to launch any communication kernels written for NCCL. Note that NCCL communication APIs cannot be called from inside a CUDA kernel. NCCL differs from NVSHMEM in that NVSHMEM provides both GPU and CPU-side APIs for its one-sided communication and collective communication. Applications that use NCCL for communication have separate kernels launched for computation and communication, and optimizations like overlapping and pipelining are achieved by using CUDA streams and synchronization barriers. Using NVSHMEM allows applications implemented in CUDA kernels to execute both computation and communication. An example [27] illustrates a multi-GPU Jacobi kernel written using both NCCL and NVSHMEM to show that one-sided communication API can be used in place of NCCL APIs to make the code look simpler.

## 3. Performance Comparison of NCCL vs NVSHMEM

We use NVSHMEM *perftests* which are shipped with the installation of the library. This provides latency and bandwidth benchmarks for various device and host-side APIs. For NCCL, we use *nccl-tests* [28], an open-source performance benchmark provided by NVIDIA for NCCL operations. We use Runpod [29], a cloud platform that provides on-demand GPUs. We run these experiments on 8 H100-SXM GPUs in a DGX configuration that is NVLink connected. While NCCL provides more primitives than NVSHMEM, we compare them across the four that are common to both. These include all reduce, all gather, all to all, and broadcast.

The all-reduce operation performs reductions (sum, max, etc.) on data across devices or ranks and writes the result in receive buffers of each rank. The all-reduce operation is one of the most widely used communication primitives in deep learning models for both training and inference. One simple example during training would be summing the gradients across devices in the case of model parallelism before the optimizer step. As shown in Figure 1, we see that for smaller message sizes up to 2KB, the execution time using NVSHMEM is about 10 times smaller than that of NCCL. For larger message sizes over 8KB, the NCCL execution time remains relatively flat and is much less than that of NVSHMEM.

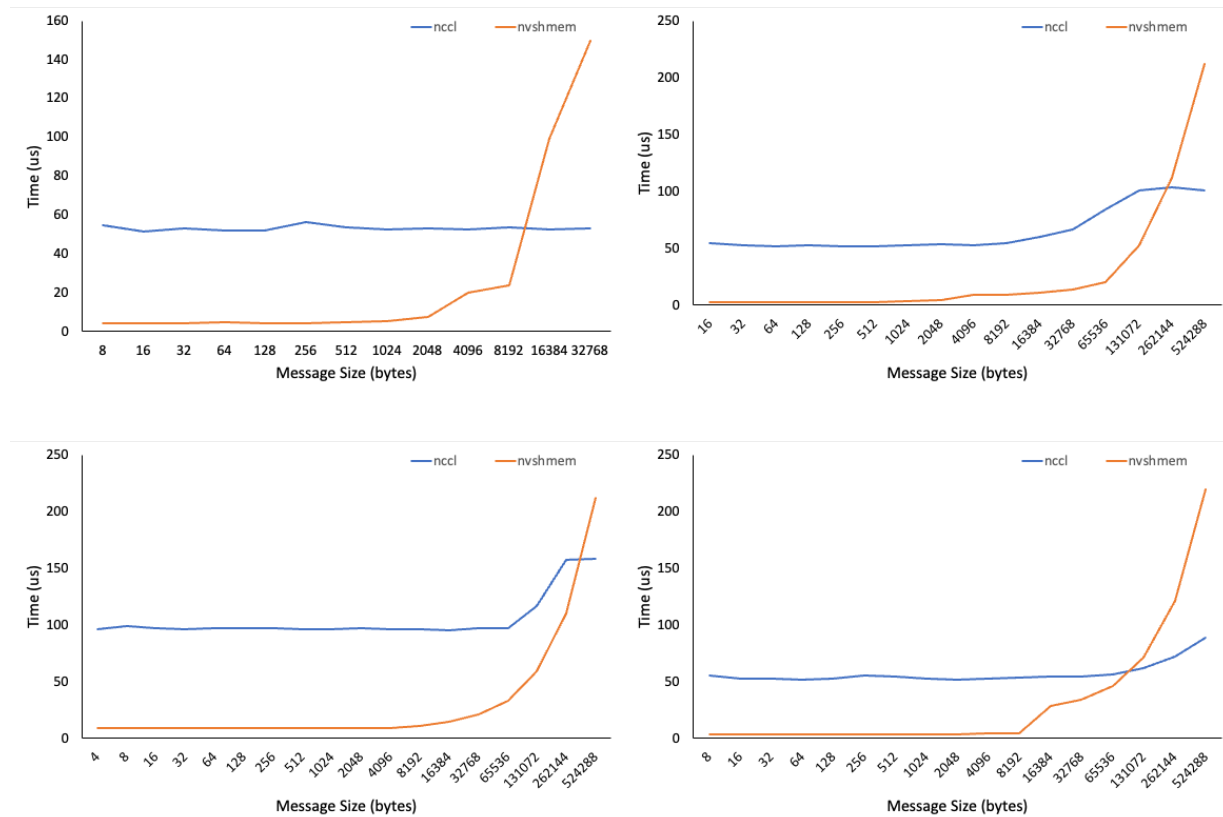

Figure 1: Execution time taken for (from top left, clockwise) i) All Reduce ii) All Gather iii) Broadcast iv) All to All across message sizes on an H100 DGX system. Blue – NCCL, Red - NVSHMEM

In an all-gather operation, each rank receives the concatenation of data from all ranks in the order of ranks. In deep-learning models, an example would be matrix multiplications in a feed-forward layer. One way to split the AB = C kind of matrix multiplication is to replicate one matrix while splitting the other by column. After the multiplication, the partial matrices are aggregated back using all gather to get the result of matrix C. From Figure 1, we see a similar trend as all reduce. In this case for smaller message sizes, the NVSHMEM execution time is about 20x shorter than that of NCCL up to 8KB. The execution times for NVSHMEM increase steadily while those of NCCL continue to remain relatively constant. In all-to-all collective communication, each device sends data to each one of the other ranks in the communication group. The all-to-all primitive is commonly used in DLRM inference when the tables are sharded row-wise. During inference, when data parallel batches are sent to each processor, every processor needs to send the indices to the GPUs where the embedding table segment resides to do the embedding rows lookup. From Figure 1, we see that NVSHMEM is about 10x faster than NCCL for smaller message sizes. The execution time for NVSHMEM increases quickly for larger message sizes and exceeds that of NCCL after 256KB. The broadcast operation sends out the data from one rank (root) to all the other ranks in the communication group. The trend follows similarly to the other communication primitives in that the execution time for NVSHMEM is significantly shorter for the smaller message sizes but is substantially longer for the larger message sizes.

## 4. Implementation of an Embedding Bag Kernel

### 4.1 DLRM Model Architecture

DLRM comprises four major sections as shown in Figure 2: a bottom-level MLP, embedding pooling, a top-level MLP, and the feature interaction for the outputs of the bottom-level MLP combined with the outputs of the embedding pooling. Embedding pooling uses an Embedding Bag to aggregate its result. An Embedding Bag functions as a lookup of one or more embedding vectors. Each of the embedding vectors acts as a latent vector encoding for a particular categorical feature value. The embedding tables encompass the collection of all embedding vectors for a particular categorical value. The large size of these embedding tables and their irregular memory access patterns make embedding bag a unique operator to optimize and shard. In this section, the focus is on the embedding pooling performance.

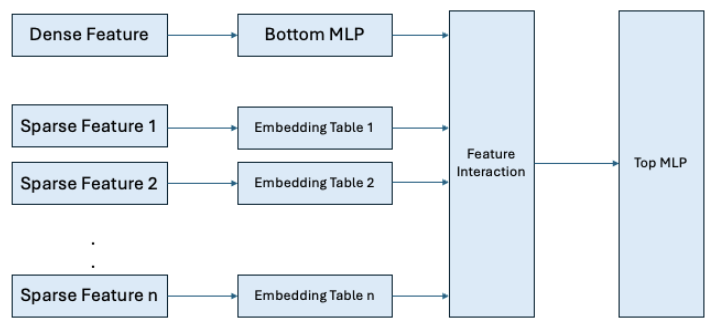

*Figure 2: Canonical Architecture of DLRM*

Training data of these models comprises both dense (numerical) features and sparse (categorical) features with most of them as sparse features. The number of sparse features determines the number of embedding tables used, and the number of rows in each of these tables is determined by distinct values for the category in the training data. The embedding dimension is the number of columns of each row, which usually falls in the range between 32 and 256. To get a sense of how big the embedding tables are in memory, an embedding table for a feature in the Criteo dataset [30] with around 40 million distinct features and an embedding dimension of 128 would amount to close to 150 GBs when the parameters are stored in fp32 data type. In a real-world scenario, we expect the total embedding size to be at least several TBs.

An Embedding Bag operator sums up different rows in the embedding table for a single inference. The pooling size for an inference refers to the number of rows that are retrieved to aggregate together for an inference. Thus, the pooling size determines the memory performance requirement of DLRM and has increased by almost 30 times in four years from 2017 [31]. The irregular and data-dependent memory access patterns of embedding rows make it a challenge for models that are split across multiple GPUs. For example, when we split one million rows table across GPUs, and inference for the model pools the rows from each of these sharded parts, an additional aggregation step is needed to combine the results from the respective GPUs to be pooled together.

The unique nature of the embedding bag operator implies that the model parallelism strategies would look different than that of other compute-intensive deep-learning models. The three main optimization techniques for embedding bag are as follows: Row Wise Parallelism (RW), Column Wise Parallelism (CW), and Table Wise Parallelism (TW) [3]. Row parallelism means the embedding tables are sliced by rows; for example, across two GPUs, half of the rows of the table are in GPU 0 and another half in GPU 1. In column-wise parallelism, the tables are split across columns, and in table-wise parallelism, some tables are placed on one GPU and others on another GPU assuming that the size of a table is less than the available memory of one GPU.

## 4.2 Using Row Wise Parallelism

When a DLRM model is sharded row-wise, rows of an embedding table are distributed across GPUs. During inference, each of these GPUs gets a mini-batch of inputs, categorical, and numerical values. The numerical values flow through to the Bottom MLP portion of the model, while the categorical inputs are indices of the rows that need to be looked up for each of the categorical features. The input format of any of these categorical features is described with two values:

i) Indices: array of indices to be looked up from the respective embedding table.
   For example, an embedding table that has 20 rows could have indices that look like *[14, 29, 12, 6, 13, 10, 8, 2]*. But this alone does not give information about the batch size or pooling size.
ii) Lengths: array of length of indices to be picked for each inference i.e. pooling size.
   For example, *[2, 1, 0, 3, 2]* lengths array along with the above input would mean the batch size is 5, and the 1$^{st}$ inference would need to sum up the embedding table rows indexed at 14 and 29. The 2$^{nd}$ inference would need only a row indexed at 12.

We will focus entirely on the embedding bag portion of the model in inference mode, in which given the inputs, the model predicts the outputs. As shown in Figure 4, each minibatch would contain indices and lengths for each of the embedding tables in the model. These input indices now need to be shuffled or permuted by every GPU, so that every index is sent to the respective GPUs to be retrieved. This would constitute an all-to-all communication because each GPU needs to send some part of its data to every other GPU based on their input indices and sharding plan. Afterward, every GPU would perform a gather operation on indices of their chunks in the embedding table. The last step would be adding back the results from each of the chunks and sending them to the original-requested GPUs for those indices in a reduce-scatter operation.

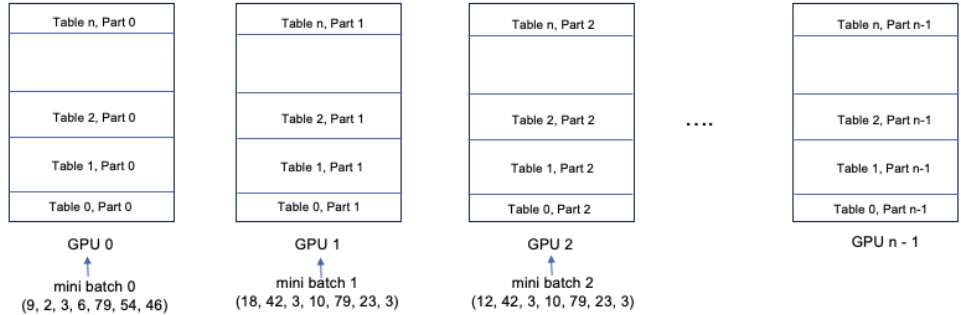

Figure 2: Row Wise Parallelism

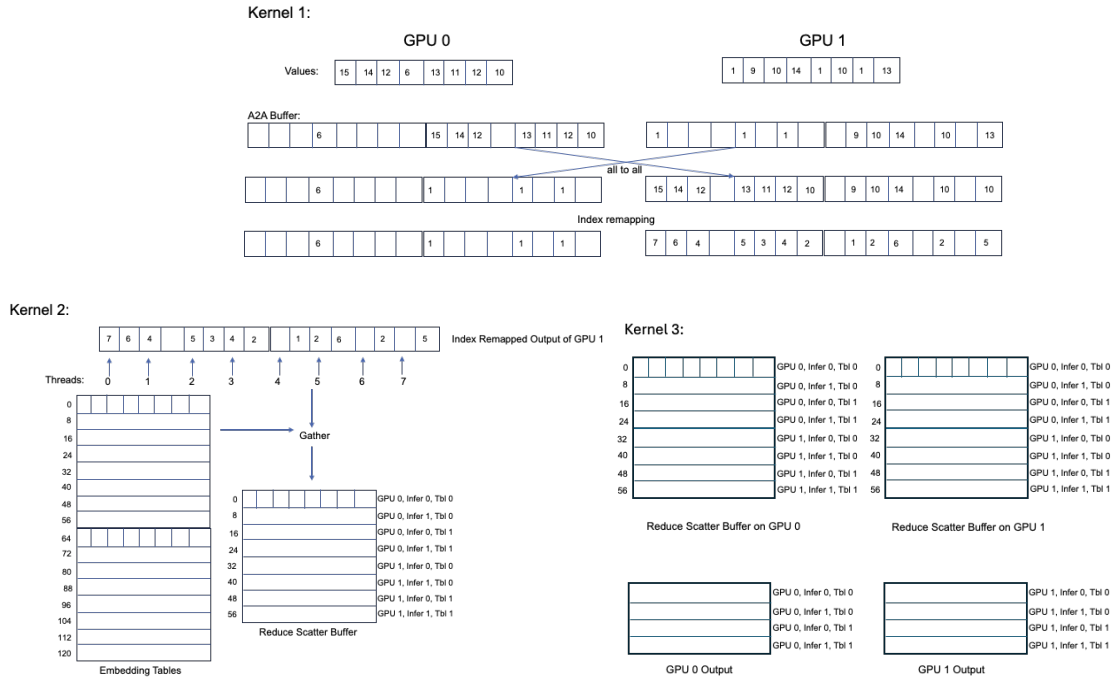

Figure 3: An example (Embedding Dimension: 8, Number of tables: 2, mini-batch size: 2, pooling: 2) of the kernels. (Top) Permute kernel (all to all) for input indices, (Bottom Left) Gather operation, (Bottom Right) Reduce Scatter for output.

## 4.3 Experimental Assumptions

We implement an example of the above-described process with NVSHMEM for the communication primitives. As shown in Figure 5, there are three major CUDA kernels to implement for this row-wise parallel Embedding Bag operator. First, a permute and an all-to-all kernel decides which embedding table is sharded for a given index so that the index can be sent to the corresponding GPU that holds the shard. Second, a gather kernel does a lookup of the embedding rows and a pooling of the lookups and stores the results in the reduce-scatter buffer. Finally, a reduce-scatter kernel is executed to get the final outputs of each mini-batch on the GPUs.

Several variables impact this operator such as the sharding plan to decide which rows to map to which GPUs, pooling factors, data type of the embedding table parameters, the number of embedding tables, rows per embedding table, and embedding dimension. We make some assumptions to simplify the implementation of CUDA kernels for embedding bag. We aim to show an implementation of a deep learning operator using a communication library for deep-learning use cases. The following assumptions are made as follows: the number of rows/parameters in each table is the same, each table is split evenly across GPUs row-wise, and the pooling size for each inference is constant for all the tables.

### 4.4 Implementation Validation

TorchRec is a PyTorch domain library that incorporates various parallelism primitives needed for the training of large-scale recommender systems in which huge embedding tables are sharded across multiple GPUs. TorchRec uses NCCL as the primary collective communication library. We validate our implementation using NVSHMEM with TorchRec by importing the embedding table weights and comparing the outputs of the same for functional correctness. We randomly initialized embedding table weights in TorchRec and randomly generated indices based on the number of rows of each embedding table.

Our NVSHMEM implementation of the Embedding Bag is compared with the TorchRec Embedding Bag that uses NCCL. We use 8 H100-SXM GPUs in a DGX configuration and use randomly generated data for inputs and randomly initialized weights to populate the embedding tables in TorchRec. The number of embedding table rows of each table is fixed in both cases, and we vary other parameters including the batch size, the number of embedding tables, and the number of pooling factors. The embedding dimension is fixed at 128 bytes per row.

The three kernels are measured by modifying the source code of the TorchRec library to time the corresponding portions of the communication and compute times. We test different configurations of the embedding bag operator by changing the number of embedding tables from 2 to 64 with increments of multiple of 2. Different batch sizes of 128, 1024, and 4096 are used along with changing the pooling factors to 4, 8, and 16 for each inference.

The results of the performance comparison between NCCL and NVSHMEM for the Embedding Bag operator are shown in Figures 6, 7, and 8. Similar to what is found for individual communication collective performance described earlier, we see that the performance of the Embedding Bag operator has shorter execution times for NVSHMEM implementation when the total message size for communication is smaller. After a certain threshold is reached and it varies for different experimental parameters, the Embedding Bag performance flips over to shorter execution times for NCCL implementation as the message sizes increase.

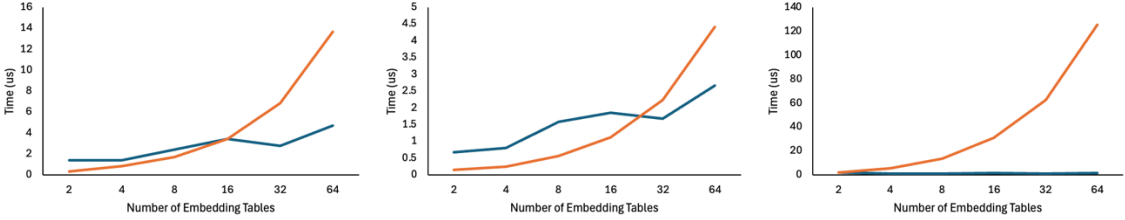

*Figure 4: Across different number of embedding tables. i) Indices Permute ii) Gather rows iii) Reduce Scatter. Red: NVSHMEM, Blue: NCCL*

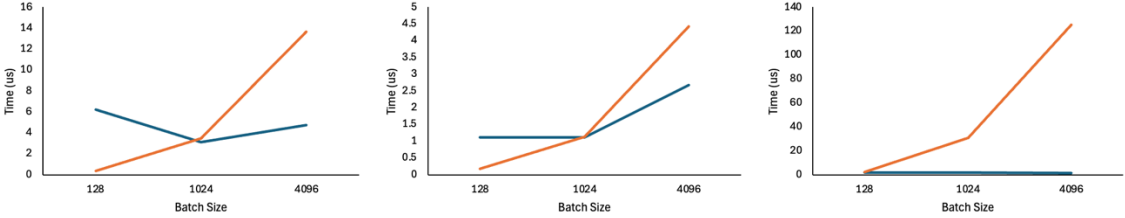

*Figure 5: Across different batch sizes. i) Indices Permute ii) Gather rows iii) Reduce Scatter. Red: NVSHMEM, Blue: NCCL*

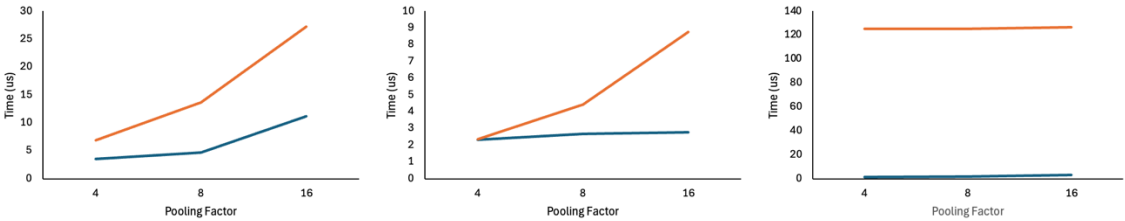

*Figure 6: Across different pooling factors. i) Indices Permute ii) Gather rows iii) Reduce Scatter. Red: NVSHMEM, Blue: NCCL*

In NVSHMEM 2.9, we do not have a communication primitive API for reduce scatter. Hence, we implemented a simple reduce scatter kernel which achieves the same result by using the NVSHMEM all-to-all API and then summing in respective GPUs.

## 5. Performance Impact of Distributing Embeddings Across GPUs

### 5.1. Experiments

In this study, we utilized the TorchRec library alongside NCCL, serving as the backend for collective communication, to assess the performance of embedding pooling, which is identified as a critical bottleneck in DLRMs. The embedding entails input redistribution, embedding pooling, and output reduce scatter. We implemented row-wise parallelism across multiple GPUs to facilitate our simulation. The experimental setup involved a single NVIDIA DGX box equipped with eight H100 GPUs interconnected using NVLink. Our tests were designed to cover various configurations. We evaluated a single embedding table with batch sizes per GPU set at 128, 256, 512, and 1024, embedding dimensions at 32, 64, 128, and 256, and pooling factors per GPU at 4, 8, and 16. In addition, we tested multiple embedding tables from 1, 2, 4, 8, 16, 32, and 64 with batch sizes per GPU of 128, 1024, and 4096, a pooling factor of 32, and embedding dimensions of 32 and 128. Both experiment sets are evaluated across 1, 2, 4, and 8 GPU configurations.

Additionally, to assess the impact of interconnect technologies, we conducted nccl-tests on an H100-DGX box using PCIe and compared these results to tests on the same system utilizing NVLink connections.

## 5.2. Results

We put all the pieces together to formulate an overall performance of DLRMs by projecting the performance of embedding pooling. Figure 9 shows the projected speedup of an embedding pooling that can be performed entirely in its locally addressable memory over that of an embedding table that requires distributing across multiple GPUs. The number of GPUs used depends on the embedding table size by dividing it by the amount of HBM memory available in a GPU (i.e., 80 GB for an H100 GPU). For example, a 10 TB embedding table would use 128 GPUs. In this case, the performance speedup is projected to be from 22.8x to 108.2x depending on the total message size as determined by the number of embedding tables performed in a batch fashion, the average number of pooling factors, and the embedding dimensions.

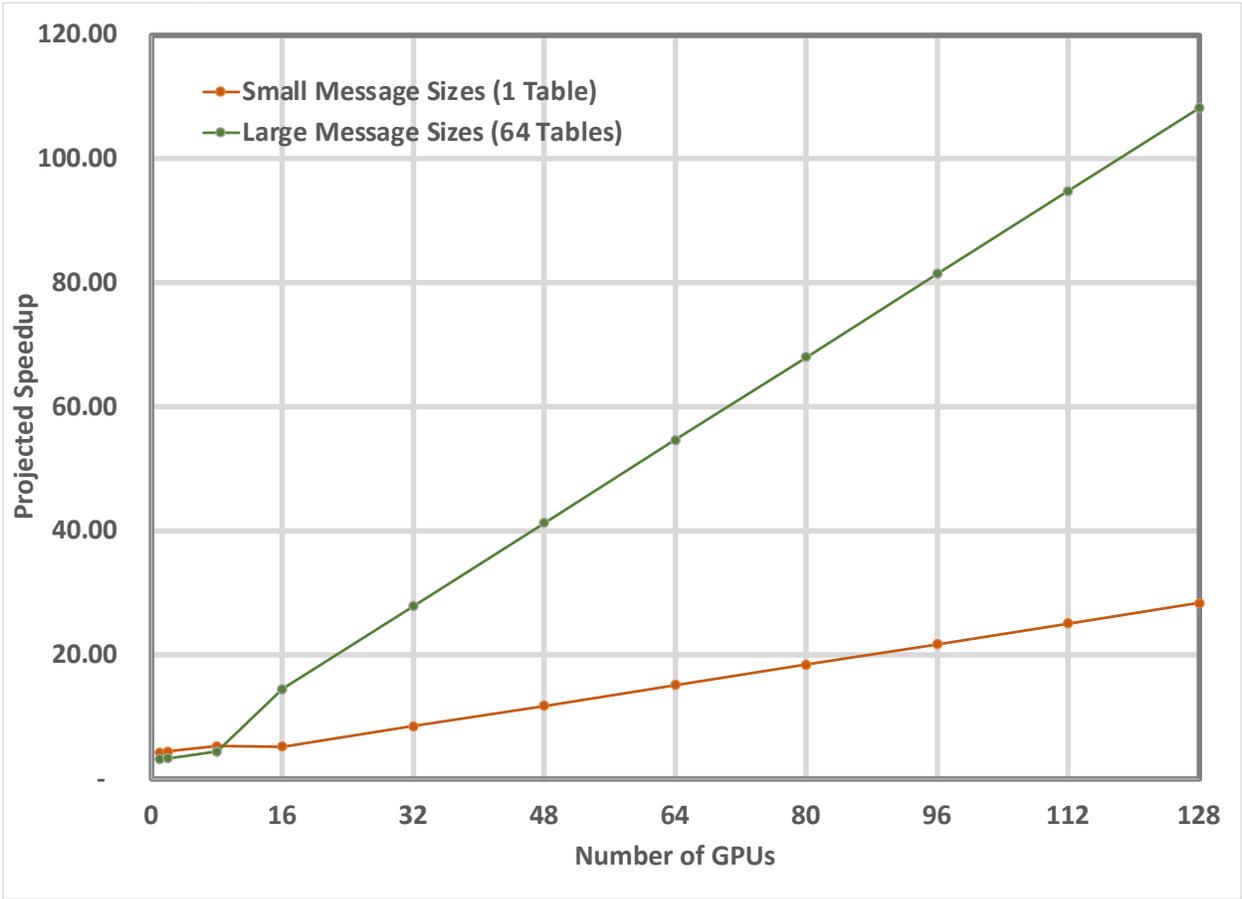

Figure 9: Performance speedup for different embedding table sizes as determined by the number of GPUs used and the amount of HBM memory in each GPU

## 6. Summary

As optimizing DLRM performance is directly tied to improving the embedding pooling performance, we compare the performance of an Embedding Bag operator implemented in both NCCL and NVSHMEM across different batch sizes, number of tables, table sizes, pooling factors, and embedding dimensions. With irregular memory accesses of the row retrievals of an embedding table, the embedding pooling is further complicated by the need for an entire embedding table to be distributed across multiple GPUs. Distributing embedding incurs a significant communication and synchronization cost of sending and receiving data among a cluster of GPUs despite many improvements in interconnect technology. We collected experimental data on H100 GPUs to project the performance speedup to be at least an order of magnitude higher for an embedding table using locally addressable memory without incurring additional overhead from distributing embedding across multiple GPUs.

References

[1] He, K., Zhang, X., Ren, S., & Sun, J. (2016). Deep residual learning for image recognition. In *Proceedings of the IEEE conference on computer vision and pattern recognition* (pp. 770-778).

[2] Naumov, M., Mudigere, D., Shi, H. J. M., Huang, J., Sundaraman, N., Park, J., ... & Smelyanskiy, M. (2019). Deep learning recommendation model for personalization and recommendation systems. *arXiv preprint arXiv:1906.00091*.

[3] Mudigere, D., Hao, Y., Huang, J., Jia, Z., Tulloch, A., Sridharan, S., ... & Rao, V. (2022, June). Software-hardware co-design for fast and scalable training of deep learning recommendation models. In *Proceedings of the 49th Annual International Symposium on Computer Architecture* (pp. 993-1011).

[4] Zha, D., Feng, L., Bhushanam, B., Choudhary, D., Nie, J., Tian, Y., ... & Hu, X. (2022, August). Autoshard: Automated embedding table sharding for recommender systems. In *Proceedings of the 28th ACM SIGKDD Conference on Knowledge Discovery and Data Mining* (pp. 4461-4471).

[5] Zha, D., Feng, L., Tan, Q., Liu, Z., Lai, K. H., Bhushanam, B., ... & Hu, X. (2022). Dreamshard: Generalizable embedding table placement for recommender systems. *Advances in Neural Information Processing Systems*, *35*, 15190-15203.

[6] Lepikhin, D., Lee, H., Xu, Y., Chen, D., Firat, O., Huang, Y., ... & Chen, Z. (2020). Gshard: Scaling giant models with conditional computation and automatic sharding. *arXiv preprint arXiv:2006.16668*.

[7] Smith, B., & Linden, G. (2017). Two decades of recommender systems at Amazon.com. *IEEE internet computing*, *21*(3), 12-18.

[8] Ma, Y., Narayanaswamy, B., Lin, H., & Ding, H. (2020, August). Temporal-contextual recommendation in real-time. In *Proceedings of the 26th ACM SIGKDD international conference on knowledge discovery & data mining* (pp. 2291-2299).

[9] Gomez-Uribe, C. A., & Hunt, N. (2015). The Netflix recommender system: Algorithms, business value, and innovation. *ACM Transactions on Management Information Systems (TMIS)*, *6*(4), 1-19.

[10] Cheng, H. T., Koc, L., Harmsen, J., Shaked, T., Chandra, T., Aradhye, H., ... & Shah, H. (2016, September). Wide & deep learning for recommender systems. In *Proceedings of the 1st workshop on deep learning for recommender systems* (pp. 7-10).

[11] Liu, Z., Zou, L., Zou, X., Wang, C., Zhang, B., Tang, D., ... & Cheng, Y. (2022). Monolith: real time recommendation system with collisionless embedding table. *arXiv preprint arXiv:2209.07663*.
[12] NVIDIA, [Online]. Available: https://www.nvidia.com/en-us/data-center/h100/.
[13] NVIDIA, [Online]. Available: https://www.nvidia.com/en-us/data-center/nvlink/.
[14] NVIDIA, [Online]. Available: https://www.nvidia.com/en-us/networking/products/infiniband/.
[15] AMD, [Online]. Available: https://www.amd.com/en/technologies/infinity-architecture.
[16] NVIDIA, [Online]. Available: https://github.com/NVIDIA/nccl.
[17] NVIDIA, [Online]. Available: https://developer.nvidia.com/nvshmem.
[18] Gupta, U., Wu, C. J., Wang, X., Naumov, M., Reagen, B., Brooks, D., ... & Zhang, X. (2020, February). The architectural implications of Facebook's dnn-based personalized recommendation. In *2020 IEEE International Symposium on High Performance Computer Architecture (HPCA)* (pp. 488-501). IEEE.
[19] Sethi, G., Bhattacharya, P., Choudhary, D., Wu, C. J., & Kozyrakis, C. (2023). FlexShard: Flexible Sharding for Industry-Scale Sequence Recommendation Models. *arXiv preprint arXiv:2301.02959*.
[20] PyTorch [Online]. Available: https://pytorch.org/.
[21] TensorFlow [Online]. Available: https://www.tensorflow.org/.
[22] OpenSHMEM [Online]. Available: http://openshmem.org/site/.
[23] D.W. Walker. Standards for message-passing in a distributed memory environment. Technical Report TM-12147, Oak Ridge National Laboratory, 1992.
[24] C. Coarfa, Y. Dotsenko, J. Mellor-Crummey, F. Cantonnet, T. ElGhazawi, A. Mohanty and Y. Yao, "An Evaluation of Global Address Space Languages: Co-Array Fortran and Unified Parallel C," Proc. of the 10th ACM SIGPLAN symp. on Principles and practice of parallel programming, pp. 36-47, 2005.
[25] B. Chapman, G. Jost, R. van der Pas, Using, OpenMP: Portable Shared Memory Parallel Programming. MIT Press, 2007.
[26] OpenCL 1.1 Specification. http://www.khronos.org/registry/cl/specs/\ opencl-1.1.pdf. Oct. 2011.
[27] NVIDIA-GitHub-multi-gpu-programming-models. [Online] https://github.com/NVIDIA/multi-gpu-programming-models.
[28] NVIDIA GitHub – nccl-tests.[Online] https://github.com/NVIDIA/nccl-tests
[29] Runpod, [Online]. Available: https://www.runpod.io/.
[30] Criteo Dataset. [Online]. Available: https://labs.criteo.com/2013/12/download-terabyte-click-logs/.
[31] Sethi, G., Acun, B., Agarwal, N., Kozyrakis, C., Trippel, C., & Wu, C. J. (2022, February). RecShard: statistical feature-based memory optimization for industry-scale neural recommendation. In *Proceedings of the 27th ACM International Conference on Architectural Support for Programming Languages and Operating Systems* (pp. 344-358).
[32] Gholami, A., Yao, Z., Kim, S., Hooper, C., Mahoney, M. W., & Keutzer, K. (2024). Ai and memory wall. *IEEE Micro*.